\documentclass[preprint,pra,aps,superscriptaddress,showpacs,prd]{revtex4}
\usepackage{epsfig,amsmath,graphicx,amssymb}
\def\be{\begin{equation}}
\def\ee{\end{equation}}
\def\bea{\begin{eqnarray}}
\def\eea{\end{eqnarray}}
\def\bes{\begin{subequations}}
\def\ees{\end{subequations}}
\begin{document}
\title{Controllable Entanglement of Lights in a Five-Level System}

\author{Yun Li}
\address{Department of Physics, East China Normal University,
Shanghai 200062, China}

\author{Chao Hang}
\address{Department of Physics, East China Normal University,
Shanghai 200062, China}

\author{Lei Ma}
\email{lma@phy.ecnu.edu.cn}
\address{Department of Physics, East China Normal University,
Shanghai 200062, China}

\author{Guoxiang Huang}
\email{gxhuang@phy.ecnu.edu.cn}
\address{Department of Physics, East China Normal University,
Shanghai 200062, China}

\date{\today}


\begin{abstract}

We analyze the nonlinear optical response of a five-level system
under a novel configuration of electro-magnetically induced
transparency. We show that a giant Kerr nonlinearity with a
relatively large cross-phase modulation coefficient that occurs in
such system may be used to produce an efficient photon-photon
entanglement. We demonstrate that such photon-photon entanglement
is practically controllable and hence facilitates promising
applications in quantum information and computation.

\end{abstract}

\maketitle

\section{Introduction}

Entanglement is one of the most profound features of quantum
mechanics\cite{EPR}. A system consisting of two subsystems is said
to be entangled if its quantum state cannot be described by a
product of the quantum states of the two
subsystems\cite{Bennett1996}. In past decades, entanglement has
been the focus of a large amount of study on the foundation of
quantum mechanics, associated particularly with quantum
nonseparability and violation of Bell's
inequalities\cite{Ballentine}. It has also been viewed as a
potential resource that can be used for many practical
applications, especially for quantum communication\cite{Pan1997},
cryptography\cite{Gisin}, and computation\cite{Williams}.

It is important for quantum information processing to be able to
create entangled states in a controllable way. In recent years,
many novel methods have been proposed to generate controllable
entangled states\cite{UnanyanA2001,UnanyanL2001,Emary}. Some of
them are based on quantum interference effects associated with
electromagnetically induced transparency (EIT)\cite{Harris},
including atom-atom, atom-photon and photon-photon entangled
states\cite{Lukin2000,Paternostro,Kuang,Payne,Kiffner,Jing,Peng,Friedler}.
Comparing with conventional nonresonant media\cite{Linares,Kiess},
atomic systems under an EIT configuration possess many striking
features, such as very low absorption, ultraslow group velocity,
and enhanced Kerr nonlinearity under weak driving
conditions\cite{fle}. It has also been  shown recently that these
properties can be used to produce a new type of optical solitons,
i.e. ultraslow optical solitons\cite{wu,hua1}.

As is well known, Kerr nonlinearity is important for producing an
interaction between light fields. It is also crucial to get an
efficient photon-photon entanglement\cite{nie}. In an EIT medium,
a giant enhancement of Kerr nonlinearity can be obtained by a
slightly disturbance of resonance condition. Up to now, different
schemes has been proposed for obtaining such enhancement through a
cross-phase modulation (CPM) effect, including ``N''
configuration\cite{Schmidt,Kang}, chain-$\Lambda$ configuration
\cite{Zubairy,Matsko,Greentree} and tripod
configuration\cite{Petrosyan}. It has also been suggested to
achieve a large nonlinear mutual phase
shift\cite{Schmidt,Lukin2000} and to construct all-optical
two-qubit quantum phase gates  under weak driving
conditions\cite{Ottaviani, Rebic}.

In this paper, we investigate the photon-photon entanglement in a
five-level atomic system under a novel EIT configuration. Our
study exhibits several important features. Firstly, the self-phase
modulation (SPM) effect can be detached and suppressed, which is
different from the case with a tripod configuration\cite{Rebic},
where the effect of SPM counteracts that of CPM and impairs the
formation of the entanglement. Secondly, an enhanced CPM effect
can be obtained in our system because the EIT configuration in our
system has a better symmetry, which makes not only the group
velocities of both the probe and trigger pulses decrease several
orders of magnitude but also the group-velocity matching can be
achieved much easier than the case using a chain-$\Lambda$
configuration\cite{Ottaviani}. In addition, in our approach the
effect of absorption and dispersion is taken into account. When
the binary information is encoded in the polarization degree of
freedom of the probe and the trigger pulses,  the entanglement
between the probe and trigger pulses relies on the large nonlinear
mutual phase shift contributed by the enhanced CPM effect. We
shall show that such photon-photon entanglement is controllable
and hence may facilitate more applications in in quantum
information and computation. The paper is arranged as follows. The
following section (Sec. II) describes the model under study. In
Sec. III linear and nonlinear susceptibilities of the system are
calculated based on Bloch equations. The group-velocity matching
and the deformation of the probe and trigger pulses due to the
dispersion and absorption effects are also discussed. In Sec. IV
we investigate the degree of entanglement of two-qubit states with
realistic parameters by using entanglement of formation. Finally,
Sec. V contains a discussion and a summary of our results.

\section{The Model}

We consider a life-time broadened five-state atomic system, shown
in Fig. 1. The system interacts with a weak, pulsed probe field of
center frequency $\omega_{P}/(2\pi)$
($|0\rangle\rightarrow|4\rangle$ transition), a weak, pulsed
trigger field $\omega_{T}/(2\pi)$ ($|2\rangle\rightarrow|3\rangle$
transition) and two strong, continuous-wave (cw) coupling fields
of frequencies $\omega_{B}/(2\pi)$
($|1\rangle\rightarrow|3\rangle$ transition) and
$\omega_{C}/(2\pi)$ ($|1\rangle\rightarrow|4\rangle$ transition).
The electric-field can be written as $E=\frac{1}{2}({\cal E}_P
\,e^{-i\omega_P t}$
        +${\cal E}_B \,e^{-i\omega_B t}$
        +${\cal E}_C \,e^{-i\omega_C t}$
        +${\cal E}_T \,e^{-i\omega_T t}$
        )+${\rm c}.{\rm c.}$, where c.c. represents complex
        conjugate.

\begin{figure}
\centering
\includegraphics[scale=0.55]{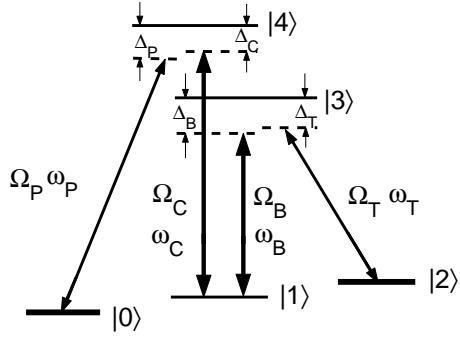}
\caption{\footnotesize The energy-level diagram and excitation
scheme of a life-time broadened five-state atomic system
interacting with two weak, pulsed (probe and trigger) fields of
frequency $\omega_{P}/(2\pi)$ and $\omega_{T}/(2\pi)$ and two
strong, cw control fields of frequencies $\omega_{B}/(2\pi)$ and
$\omega_{C}/(2\pi)$.}
\end{figure}

In Schr\"{o}dinger picture the Hamiltonian of the system is given
by $H=H_0+H_1$,  with
\bea
& H_0=&\hbar\omega_0 |0\rangle\langle0|+\hbar\omega_1|1\rangle
\langle1|+\hbar\omega_2 |2\rangle\langle2| +\hbar\omega_3
|3\rangle\langle3|+\hbar\omega_4 |4\rangle\langle4|\\
&H_1=&-\hbar(\Omega_{P}^{*}e^{i\omega_Pt}|0\rangle \langle4|
   +\Omega_{B}^{*}e^{i\omega_Bt}|1\rangle\langle3|
   +\Omega_{C}^{*}e^{i\omega_Pt}|1\rangle\langle4|
   \nonumber\\
& &+\Omega_{T}^{*}e^{i\omega_Pt}|2\rangle\langle3|+
   {\rm h}.{\rm c}.),
\eea
where $\Omega_{l}$  ($l$=$P,B,C,T$) are the half Rabi frequencies
of relevant fields with $\Omega_{P}=D_{40}{\cal E}_P/(2\hbar)$,
$\Omega_{T}=D_{32}{\cal E}_T/(2\hbar)$, $\Omega_{B}=D_{31}{\cal
E}_B/(2\hbar)$ and $\Omega_{C}=D_{41}{\cal E}_C/(2\hbar)$.
$D_{ij}=D_{ji}^{*}$ are the electric dipole matrix elements.
Within a rotating-wave approximation the evolution of the system
is described by the Bloch equations for density-matrix elements
\bes \label{Bloch}
\bea
& &\dot{\sigma}_{00}=-\gamma_{00}\sigma_{00}+i\Omega_{P}^*
   \sigma_{40}-i\Omega_{P}\sigma_{04},\\
& &\dot{\sigma}_{11}=-\gamma_{11}\sigma_{11}+i\Omega_{B}^*
   \sigma_{31}-i\Omega_{B}\sigma_{13}+i\Omega_{C}^*\sigma_{41}
   -i\Omega_{C}\sigma_{14},\\
& &\dot{\sigma}_{22}=-\gamma_{22}\sigma_{22}+i\Omega_{T}^*
   \sigma_{32}-i\Omega_{T}\sigma_{23},\\
& &\dot{\sigma}_{33}=-\gamma_{33}\sigma_{33}+i\Omega_{B}
   \sigma_{13}-i\Omega_{B}^*\sigma_{31}+i\Omega_{T}\sigma_{23}
   -i\Omega_{T}^*\sigma_{32},\\
& &\dot{\sigma}_{44}=-\gamma_{44}\sigma_{44}+i\Omega_{P}
   \sigma_{04}-i\Omega_{P}^*\sigma_{40}+i\Omega_{C}\sigma_{14}
   -i\Omega_{C}^*\sigma_{41},\\
& &\dot{\sigma}_{01}=-(\gamma_{01}+i\Delta_C-i\Delta_P)
   \sigma_{01}+i\Omega_P^*\sigma_{41}-i\Omega_B\sigma_{03}
   -i\Omega_C\sigma_{04},\\
& &\dot{\sigma}_{02}=-(\gamma_{02}+i\Delta_C+i\Delta_T
   -i\Delta_P-i\Delta_B)\sigma_{02}+i\Omega_{P}^*\sigma_{42}
   -i\Omega_T\sigma_{03},\\
& &\dot{\sigma}_{03}=-(\gamma_{03}+i\Delta_C-i\Delta_P
   -i\Delta_B)\sigma_{03}+i\Omega_P^*\sigma_{43}-i\Omega_B^*
   \sigma_{01}-i\Omega_T^*\sigma_{02},\\
& &\dot{\sigma}_{04}=-(\gamma_{04}-i\Delta_P)\sigma_{04}
   +i\Omega_P^*(\sigma_{44}-\sigma_{00})-i\Omega_C^*\sigma_{01},\\
& &\dot{\sigma}_{12}=-(\gamma_{12}+i\Delta_T-i\Delta_B)
   \sigma_{12}+i\Omega_B^*\sigma_{32}+i\Omega_C^*\sigma_{42}
   -i\Omega_T\sigma_{13},\\
& &\dot{\sigma}_{13}=-(\gamma_{13}-i\Delta_B)\sigma_{13}
   +i\Omega_B^*(\sigma_{33}-\sigma_{11})+i\Omega_C^*\sigma_{43}
   -i\Omega_T^*\sigma_{12},\\
& &\dot{\sigma}_{14}=-(\gamma_{14}-i\Delta_C)\sigma_{14}
   +i\Omega_B^*\sigma_{34}+i\Omega_C^*(\sigma_{44}-\sigma_{11})
   -i\Omega_P^*\sigma_{10},\\
& &\dot{\sigma}_{23}=-(\gamma_{23}-i\Delta_T)\sigma_{23}
   +i\Omega_T^*(\sigma_{33}-\sigma_{22})-i\Omega_B^*\sigma_{21},\\
& &\dot{\sigma}_{24}=-(\gamma_{24}+i\Delta_B-i\Delta_C-i\Delta_T)
   \sigma_{24}+i\Omega_T^*\sigma_{34}-i\Omega_P^*\sigma_{20}
   -i\Omega_C^*\sigma_{21},\\
& &\dot{\sigma}_{34}=-(\gamma_{34}+i\Delta_B-i\Delta_C)\sigma_{34}
   +i\Omega_B\sigma_{14}+i\Omega_T\sigma_{24}-i\Omega_P^*\sigma_{30}
   -i\Omega_C^*\sigma_{31},
\eea
\ees
where $\sigma_{ij}=\rho_{ij}\exp(-i\omega_{ij}t)$ ($i, j$=0 to 4).
$\gamma_{ij}$ describe decay of populations (i=j) and coherences
($i\neq j$). $\Delta_{P}=\omega_{40}-\omega_{P}$,
$\Delta_{B}=\omega_{31}-\omega_{B}$,
$\Delta_{C}=\omega_{41}-\omega_{C}$ and
$\Delta_{T}=\omega_{32}-\omega_{T}$ are frequency detunings.

\section{large cross-Kerr nonlinearity and group-velocity matching}

For solving the Bloch Eq. (\ref{Bloch}) we assume that the
temporal duration of the probe and trigger fields is longer enough
so that a steady state approximation can be
employed\cite{Ottaviani,Rebic}. When the intensity of the probe
and trigger fields is much weaker than the intensity of both
coupling fields, i.e. $|\Omega_{P}|^2, |\Omega_{T}|^2\ll
|\Omega_{B}|^2, |\Omega_{C}|^2$, the population in the ground
states $|0\rangle$ and $|2\rangle$ is not depleted and symmetric
with respect to $0\leftrightarrow 2$ exchange, thus
$\sigma_{00}\approx\sigma_{22}\approx 1/2$ with the population of
other three levels vanishing, i.e.
$\sigma_{11}\approx\sigma_{33}\approx\sigma_{44}\approx 0$.
Furthermore, $\sigma_{13}$, $\sigma_{14}$ and $\sigma_{34}$ vanish
 also due to small population. We solve Eq.
(\ref{Bloch}) under these consideration and  obtain the following
expressions for the susceptibilities of the probe and trigger
fields
\bes
\bea
\chi_P=n_{a}|D_{04}|^{2}\sigma_{40}/(2\epsilon_{0}\hbar\Omega_{P})=
\chi_{P}^{(1)}+\chi_{SP}^{(3)}|E_P|^2+\chi_{XP}^{(3)}|E_T|^2,\\
\chi_T=n_{a}|D_{23}|^{2}\sigma_{32}/(2\epsilon_{0}\hbar\Omega_{T})=
\chi_{T}^{(1)}+\chi_{ST}^{(3)}|E_T|^2+\chi_{XT}^{(3)}|E_P|^2,
\eea
\ees
where $\chi_{P}^{(1)}$ and $\chi_{T}^{(1)}$ are respectively the
linear susceptibilities of the probe and trigger fields, and
$\chi_{SP}^{(3)}$ and $\chi_{ST}^{(3)}$ ($\chi_{XP}^{(3)}$ and
$\chi_{XT}^{(3)}$) are respectively the nonlinear susceptibilities
characterizing the effect of self-Kerr (cross-Kerr) nonlinearity.
Since we are interested in getting a large  interaction between
the probe and trigger fields that favors the generation of
photon-photon entanglement (discussed in Sec. IV below), it is
necessary to have a relatively large cross-Kerr nonlinearity,
which can be realized in our system if one has
$|\Omega_B|^2\approx\Delta_{T}( \Delta_{T}-\Delta_{B})$ and
$|\Omega_C|^2\approx\Delta_{P}( \Delta_{P}-\Delta_{C})$. Under
these conditions\cite{note1}  the susceptibilities  related to SPM
effect, $\chi_{SP}^{(3)}$ and $\chi_{ST}^{(3)}$,  are suppressed
and can thus be neglected. Thus, we have $\chi_P\approx
\chi_{P}^{(1)}$+$\chi_{XP}^{(3)}|E_T|^2$ and $\chi_T\approx
\chi_{T}^{(1)}$+$\chi_{XT}^{(3)}|E_P|^2$, with
\bes \label{Chi}
\bea
\label{chia} & \chi_P^{(1)}=&
\frac{n_a|D_{04}|^2}{2\epsilon_0\hbar}\frac{
  2|\Omega_B|^2 -2(\Delta_P-\Delta_C)(\Delta_P+\Delta_B-\Delta_C)+i\gamma(\Delta_P-\Delta_C)}
  {N_1^{\ast}},\\
&
\chi_{XP}^{(3)}=&\frac{n_a|D_{04}|^2|D_{23}|^2}{2\epsilon_0\hbar^3}
\frac{4|\Omega_B|^2|\Omega_C|^2+(\Delta_C-\Delta_P)N_2^{\ast}}{\left(\Delta_{T}
   +\Delta_{C}-\Delta_{P}-\Delta_{B}\right)N_1^{\ast}N_2^{\ast}},\label{chib}\\
& \chi_T^{(1)}=&\frac{n_a|D_{23}|^2}{2\epsilon_0\hbar}\frac{
  2|\Omega_C|^2 -2(\Delta_T-\Delta_B)(\Delta_T+\Delta_C-\Delta_B)+i\gamma(\Delta_T-\Delta_B)}
  {N_2},\label{chic}\\
\label{chid} &
\chi_{XT}^{(3)}=&\frac{n_a|D_{23}|^2|D_{04}|^2}{2\epsilon_0\hbar^3}
\frac{4|\Omega_C|^2|\Omega_B|^2+(\Delta_B-\Delta_T)N_1}{\left(\Delta_{P}
   +\Delta_{B}-\Delta_{T}-\Delta_{C}\right)N_1N_2},
\eea
\ees
where $n_{a}$ is the atomic density, $N_1=4|\Omega_{B}|^2\Delta_P+
4[|\Omega_C|^2-\Delta_P(\Delta_P-\Delta_C)](\Delta_P+\Delta_B-\Delta_C)+\gamma^2(\Delta_P-\Delta_C)
+2i\gamma[|\Omega_B|^2+|\Omega_C|^2-(\Delta_P-\Delta_C)(2\Delta_P+\Delta_B-\Delta_C)]$
and $N_2=4|\Omega_{C}|^2\Delta_T+
4[|\Omega_B|^2-\Delta_T(\Delta_T-\Delta_B)](\Delta_T+\Delta_C-\Delta_B)+\gamma^2(\Delta_T-\Delta_B)
-2i\gamma[|\Omega_C|^2+|\Omega_B|^2-(\Delta_T-\Delta_B)(2\Delta_T+\Delta_C-\Delta_B)]$.
In above derivation, we have assumed $\Omega_P$ and $\Omega_T$ are
sufficiently weak, i.e.  $|\Omega_{P}|^2$ and $|\Omega_{T}|^2\ll
|\Omega_{B}|^2, |\Omega_{C}|^2$ and the decay rates
$\gamma_{ii}\simeq0$ (i=0 to 2), $\gamma_{33}=\gamma_{44}=\gamma$,
$\gamma_{ij}=\gamma_{34}\simeq0$ (i, j=0 to 2, $i\neq j$) and
$\gamma_{i3}=\gamma_{j4}=0.5\gamma$ (i, j=0 to 2). Due to the
symmetry of our system configuration, Eqs. (\ref{chia}),
(\ref{chib}) and Eqs. (\ref{chic}), (\ref{chid}) are symmetric
under the exchange $P\leftrightarrow T$, $B\leftrightarrow C$ and
$D_{04}\leftrightarrow D_{23}$. We stress that the imaginary parts
of the linear and nonlinear susceptibilities given above are much
smaller than their relevant real parts under the (EIT) conditions
$|\Omega_{P}|^2$ and $|\Omega_{T}|^2\ll |\Omega_{B}|^2,
|\Omega_{C}|^2$,  which result in  quantum interferences between
the states $|0\rangle$ and $|1\rangle$ and  the states $|1\rangle$
and $|2\rangle$, making the population in the states $|2\rangle$
and $|3\rangle$ be small thus very low absorption for the probe
and trigger fields\cite{note1}.

In addition to a relatively large cross-Kerr nonlinearity, group
velocity matching is another necessary condition for achieving a
large mutual phase shift because only in this way can the probe
and trigger optical pulses interact for a sufficiently long
time\cite{Lukin2000,Lukin2001}. Note that the group velocity of an
optical pulse is given by $v_{g}=c/(n_0+\omega\partial
n_0/\partial\omega)$, where $n_0=\sqrt{1+\chi^{(1)}(\omega)}$ is
linear index of refraction. Using the expressions of linear
susceptibilities given in Eq. (\ref{Chi}), we obtain
\bes \label{Goup}
\bea
&v_{g}^{P}=&\frac{1}{\dfrac{1}{c}+\dfrac{n_{a}|D_{04}|^2
\omega_P}{2\epsilon_{0}\hbar c\left(|\Omega_C|^2+|\Omega_B|^2
\right)}\left[\dfrac{1}{4}-\dfrac{|\Omega_B|^2}{4(|\Omega_B|^2+
|\Omega_C|^2)}
+\dfrac{|\Omega_B|^2}{(2\Delta-i\gamma)^2}\right]},\\
&v_{g}^{T}=&\frac{1}{\dfrac{1}{c}+\dfrac{n_{a}|D_{23}|^2
\omega_T}{2\epsilon_{0}\hbar c \left(|\Omega_C|^2+|\Omega_B|^2
\right)}\left[\dfrac{1}{4}-\dfrac{|\Omega_C|^2}{4(|\Omega_B|^2+
|\Omega_C|^2)}+\dfrac{|\Omega_C|^2}{(2\Delta-i\gamma)^2}\right]},
\eea
\ees
for the probe and trigger pulses, respectively. Actually, the
group velocities are the real part of $v_{g}^{P}$ and $v_{g}^{T}$,
which are denoted by $\tilde{v}_{g}^{P}$ and $\tilde{v}_{g}^{T}$,
used in the next section. The imaginary parts of the group
velocities result also in a damping for wave propagation. For
obtaining the above relatively simple expressions, we have assumed
that all frequency detunings are nearly, but not exactly, equal
($\simeq\Delta$). By Eq. (\ref{Goup}) the group velocity matching
can be achieved under the condition $\Omega_B\approx\Omega_C$.

Although under the EIT configuration of Fig. 1  the absorption can
be made very small but it is not vanishing and its presence may
result in an attenuation for the propagation of the probe and
trigger pulses. In addition, the dispersion effect existing in the
system will also result in a distortion of the probe and trigger
fields. For example, for a Gaussian input of the probe pulse with
the form $\Omega_P(0,0)\exp(-t^2/\tau_P^2)$, the initial amplitude
$\Omega_P(0,0)$ decreases to
$\Omega_P(0,0)/\sqrt{1-i2zG_P/\tau_P^2}$ while the initial
duration $\tau_P$ increases to $\tau_P\sqrt{1-i2zG_P/\tau_P^2}$,
where $z$ means the distance of the pulse passing through the
medium, $G_P$ denotes the group velocity dispersion (GVD) of the
probe field. The same analysis applies for the trigger field by
just taking $P\rightarrow T$\cite{note2}. The GVD of the probe and
trigger fields can be obtained by the expressions of linear
susceptibilities given in Eqs. (\ref{Chi}), which read
\bes \label{GVD}
\bea
&G_P=&\frac{n_{a}|D_{04}|^2\omega_P}{\epsilon_{0}\hbar c \left(
|\Omega_B|^2+|\Omega_C|^2 \right)}
\left[-\frac{|\Omega_C|^2(2\Delta-i\gamma)}{8(|\Omega_B|^2+
|\Omega_C|^2)^2}
+\frac{2|\Omega_B|^2}{(2\Delta-i\gamma)^3}\right],\\
&G_T=&\frac{n_{a}|D_{23}|^2\omega_T}{\epsilon_{0}\hbar c \left(
|\Omega_C|^2+|\Omega_B|^2 \right)}
\left[-\frac{|\Omega_B|^2(2\Delta-i\gamma)}{8(|\Omega_C|^2+
|\Omega_B|^2)^2}+\frac{2|\Omega_C|^2}{(2\Delta-i\gamma)^3}\right].
\eea
\ees
%

\section{Controllable Entanglement between  probe and trigger lights}

A significant interaction is key ingredient for the realization of
the entanglement between the probe and trigger fields. In our
system, such interaction can be realized by the giant CPM effect,
discussed above, by which an optical field acquires a large phase
shift conditional to the state of another optical field. We choose
two orthogonal light polarizations $|\sigma^{-}\rangle$ and
$|\sigma^{+}\rangle$ as the basis of the entanglement state.
Assuming that the input probe and trigger polarized single photon
wave packets can be expressed as a superposition of the circularly
polarized states\cite{Ottaviani}, i.e.
\begin{equation}
|\psi_i\rangle=\frac{1}{\sqrt{2}}|\sigma^-\rangle_i+\frac{1}{\sqrt{2}}
|\sigma^+\rangle_i, \quad\quad i=\{P, T\}
\end{equation}
where $|\sigma^\pm\rangle_i=\int d\omega \xi_i(\omega)
a_\pm^\dagger(\omega)|0\rangle$, with $\xi_i(\omega)$ being a
Gaussian frequency distribution of incident wave packets, centered
at frequency $\omega_i$.  The input-output relations
$|\alpha\rangle_{P}|\beta\rangle_{T} \rightarrow
\exp(i\phi_{\alpha\beta})|\alpha\rangle_{P}|\beta\rangle_{T}$  can
be satisfied. Here $\alpha,\beta=0,1$ denote two-qubit basis.

We assume the five-state system shown Fig. 1 is implemented only
when the probe has $\sigma^{+}$ polarization and the trigger has
$\sigma^{-}$ polarization. When either (both) the probe or (and)
the trigger polarizations are changed, the phase shifts acquired
by  two pulses do not involve the nonlinear susceptibilities and
lead to a difference. In fact, when both of them have the
``wrong'' polarization (probe $\sigma^{-}$ polarized and trigger
$\sigma^{+}$ polarized) there is no sufficiently close level which
the atoms can be driven to and the field acquires the trivial
vacuum phase shift $\phi_{0}^{j}=k_{j}L$ ($j=P, T$), where L is
the length of the medium. Instead, when only one of them have the
wrong polarization, the right one acquires a linear phase shift
$\phi_{0}^{j}+\phi_{lin}^{j}$, where $\phi_{lin}^{j}=2\pi
k_{j}L\text{Re}(\chi_{j}^{(1)})$. Actually, the imaginary parts of
the linear susceptibilities result in an effect of linear
absorption, which makes the output states decoherent. When none of
them have the wrong polarization, each pulse acquires a nonlinear
phase shift $\phi_{0}^{j}+\phi_{lin}^{j}+\phi_{nlin}^{j}$, where
$\phi_{nlin}^{j}$ denotes the probe or trigger nonlinear phase
shift when the five-state configuration is realized. For a
Gaussian trigger pulse of time duration $\tau_{T}$ and Rabi
frequency $\Omega_{T}$, moving with group velocity
$\tilde{v}_{g}^{T}$, the shift of the probe field reads
\be \label{phiP}
\phi_{nlin}^{P}=k_{P}\frac{\pi\hbar^{2}|\Omega_{T}|^2}{2|D_{23}|^{2}}\text{Re}(\chi_{XP}^{(3)})
\int_{0}^{L}\frac{1}{\beta_T^2}\exp\left(-\frac{\omega_Tz}{c}\text{Im}(\chi_T^{(1)})
-\left[\frac{(1/\tilde{v}_g^T-1/\tilde{v}_g^P)\sqrt{2}z}{\tau_T\beta_T}\right]^2\right)dz,
\ee
where $\beta_T=\sqrt{1-i2zG_T/\tau_T^2}$. The phase shift of the
trigger field can be obtained upon interchanging $P\leftrightarrow
T$ and $D_{04}\leftrightarrow D_{23}$ in Eq. (\ref{phiP}):
\be \label{phiT}
\phi_{nlin}^{T}=k_{T}\frac{\pi\hbar^{2}|\Omega_{P}|^2}{2|D_{04}|^{2}}\text{Re}(\chi_{XT}^{(3)})
\int_{0}^{L}\frac{1}{\beta_P^2}\exp\left(-\frac{\omega_Pz}{c}\text{Im}(\chi_P^{(1)})
-\left[\frac{(1/\tilde{v}_g^P-1/\tilde{v}_g^T)\sqrt{2}z}{\tau_P\beta_P}\right]^2\right)dz
\ee
with $\beta_P=\sqrt{1-i2zG_P/\tau_P^2}$. We should point out that
the imaginary parts of the nonlinear susceptibilities may result
in a nonlinear absorption. However, this absorption is much weaker
than the linear one and thus can be neglected. Note that the GVD
effect, reflected by the parameters $G_P$ and $G_P$, has been
taken into account in the formulas of the phase shifts
(\ref{phiP}) and (\ref{phiT}).

Using the results given above the the density matrix of the output
state is expressed as
\begin{equation*}\label{Density}
\hat{\rho}=\frac{\, 1}{Z}
\begin{pmatrix}
e^{i(\phi_{lin}^P-\phi_{lin}^{P*})} &
e^{i(\phi_{lin}^P-\phi_{lin}^{P*}-\phi_{lin}^{T*})} &
e^{i(\phi_{lin}^P+\phi_{nlin}^P+\phi_{nlin}^T)}&
e^{i(\phi_{lin}^P-\phi_{lin}^{T*})}\\
e^{i(\phi_{lin}^P-\phi_{lin}^{P*}+\phi_{lin}^T)} &
e^{i(\phi_{lin}^P-\phi_{lin}^{P*}+\phi_{lin}^T-\phi_{lin}^{T*})} &
e^{i(\phi_{lin}^P+\phi_{nlin}^P+\phi_{lin}^T+\phi_{nlin}^T)}&
e^{i(\phi_{lin}^P+\phi_{lin}^T-\phi_{lin}^{T*})}\\
e^{-i(\phi_{lin}^{P*}+\phi_{nlin}^P+\phi_{nlin}^T)}&
e^{-i(\phi_{lin}^{P*}+\phi_{nlin}^P+\phi_{lin}^{T*}+\phi_{nlin}^T)}&
1&
e^{-i(\phi_{nlin}^P+\phi_{lin}^{T*}+\phi_{nlin}^T)}\\
e^{-i(\phi_{lin}^{P*}-\phi_{lin}^T)}&
e^{-i(\phi_{lin}^{P*}+\phi_{lin}^{T*}-\phi_{lin}^T)}&
e^{i(\phi_{nlin}^P+\phi_{lin}^T+\phi_{nlin}^T)}&
e^{i(\phi_{lin}^T-\phi_{lin}^{T*})}
\end{pmatrix}
\end{equation*}
in the computational basis
$\{|\sigma^{-}\rangle_{P}|\sigma^{-}\rangle_{T},
|\sigma^{-}\rangle_{P}|\sigma^{+}\rangle_{T},
|\sigma^{+}\rangle_{P}|\sigma^{-}\rangle_{T},
|\sigma^{+}\rangle_{P}|\sigma^{+}\rangle_{T}\}$, where
$Z=[1+e^{i(\phi_{lin}^P-\phi_{lin}^{P*})}]\cdot
[1+e^{i(\phi_{lin}^T-\phi_{lin}^{T*})}]$ is a normalized
coefficient. Actually, the linear phase shift counts no
contribution on the entanglement of the probe and trigger lights
because when $\phi_{nlin}^P=\phi_{nlin}^T=0$, the space spanned by
the two qubits can be expressed as a product $\rho_P\otimes\rho_T$
of pure state of its parts.

There is a variety of measures known for quantifying the degree of
entanglement in a bipartite system, including the entanglement of
distillation\cite{Bennett1996}, the relative entropy of
entanglement\cite{Vedral}, the entanglement of
formation\cite{Bennett1996} and the entanglement
witnesses\cite{Fernando}. Here, we use the entanglement of
formation as the measure of the purity and degree of entanglement
of our two-qubit state. For an arbitrary two-qubit system, it is
given by\cite{Wootters}
\begin{equation}\label{Entan}
E_F(C)=h\left( \frac{1+\sqrt{1-C^2}}{2}\right),
\end{equation}
where $h(x)=-x{\rm log}_2(x)-(1-x){\rm log}_2(1-x)$ is Shannon's
entropy function, and $C$, called ``concurrence'', is given by
\cite{Wootters}
\begin{equation}
C(\hat{\rho})={\rm max}\{0,
\lambda_1-\lambda_2-\lambda_3-\lambda_4\}
\end{equation}
with the $\lambda_i$s being the square roots of the eigenvalues,
in a decreasing order, of the Hermitian matrix
$\hat{\rho}\widetilde{\hat{\rho}}=\hat{\rho}
\hat{\sigma}_y^P\otimes\hat{\sigma}_y^T\hat{\rho}^*
\hat{\sigma}_y^P\otimes\hat{\sigma}_y^T$, here $\hat{\rho}^*$
denotes the complex conjugation of $\hat{\rho}$ in the
computational basis, and $\hat{\sigma}_y$ is the $y$-component of
the Pauli matrix. Since $E_F(C)$ is a monotonic function of $C$,
thus we can also use the concurrence directly as our measure of
entanglement.

Now we consider a typical system working with alkali atoms to
estimate the physical parameters  and the degree of the
entanglement. We take $\gamma=10^{6}$ s$^{-1}$. The detunings are
chosen as  $\Delta_{P}=80.2\gamma$, $\Delta_{T}=80.1\gamma$,
$\Delta_{B}=79.9\gamma$ and $\Delta_{C}=80.0\gamma$. The Rabi
frequencies are taken as $\Omega_{P}=0.6\gamma$,
$\Omega_{T}=0.8\gamma$, $\Omega_{B}=3.6\gamma$ and
$\Omega_{C}=4.1\gamma$. The density of the atomic gas $n_a=1.0
\times10^{13} \text{cm}^{-3}$ and the length of the medium $l=0.1$
mm. The probe and trigger have a mean amplitude of about one
photon when the beams are tightly focused and has a time duration
about five microseconds. With the above parameters, we get
$\phi_{lin}^P=-23.42+0.68i$, $\phi_{lin}^T=12.00+0.27i$,
$\phi_{nlin}^P=0.26$ and $\phi_{nlin}^T=0.05$. Note that in this
case $|\text{Re}(\chi_{XP}^{(3)})/\text{Re}(
\chi_{P}^{(1)})|\simeq 40.0$,
$|\text{Re}(\chi_{XT}^{(3)})/\text{Re}( \chi_{T}^{(1)})|\simeq
56.6$. The group velocities of the probe and trigger fields read
$v_g^P=9.7$\,\,m/s and $v_g^T=18.9$\,\,m/s, very small indeed in
comparison with the light speed in vacuum. Using these results we
obtain the degree of entanglement $E_F=3.5\%$.

Because there are many physical parameters which can be adjusted
in a fairly large extent, the entanglement of the probe and
trigger fields in our system is practically controllable.  For
example, we choose the detunning as $\Delta_{P}=40.2\gamma$,
$\Delta_{T}=40.1\gamma$, $\Delta_{B}=39.9\gamma$ and
$\Delta_{C}=40.0\gamma$ and the Rabi frequencies as
$\Omega_{P}=0.6\gamma$, $\Omega_{T}=0.8\gamma$,
$\Omega_{B}=2.5\gamma$ and $\Omega_{C}=2.7\gamma$. The density of
the atomic gas is retained as above but the length of the medium
$l=0.03$mm. By a similar calculation we obtain
$\phi_{lin}^P=-7.24+0.65i$, $\phi_{lin}^T=4.13+0.32i$,
$\phi_{nlin}^P=0.36$ and $\phi_{nlin}^T=0.08$ (where
$|\text{Re}(\chi_{XP}^{(3)})/\text{Re}( \chi_{P}^{(1)})|\simeq
148.6$, $|\text{Re}(\chi_{XT}^{(3)})/\text{Re}(
\chi_{T}^{(1)})|\simeq 210.2$). We get the group velocities
$v_g^P=4.6$\,\,m/s and $v_g^T=8.1$\,\,m/s. The degree of
entanglement in this case is given by $E_F=6.3\%$.

\begin{figure}
\centering
\includegraphics[scale=0.4]{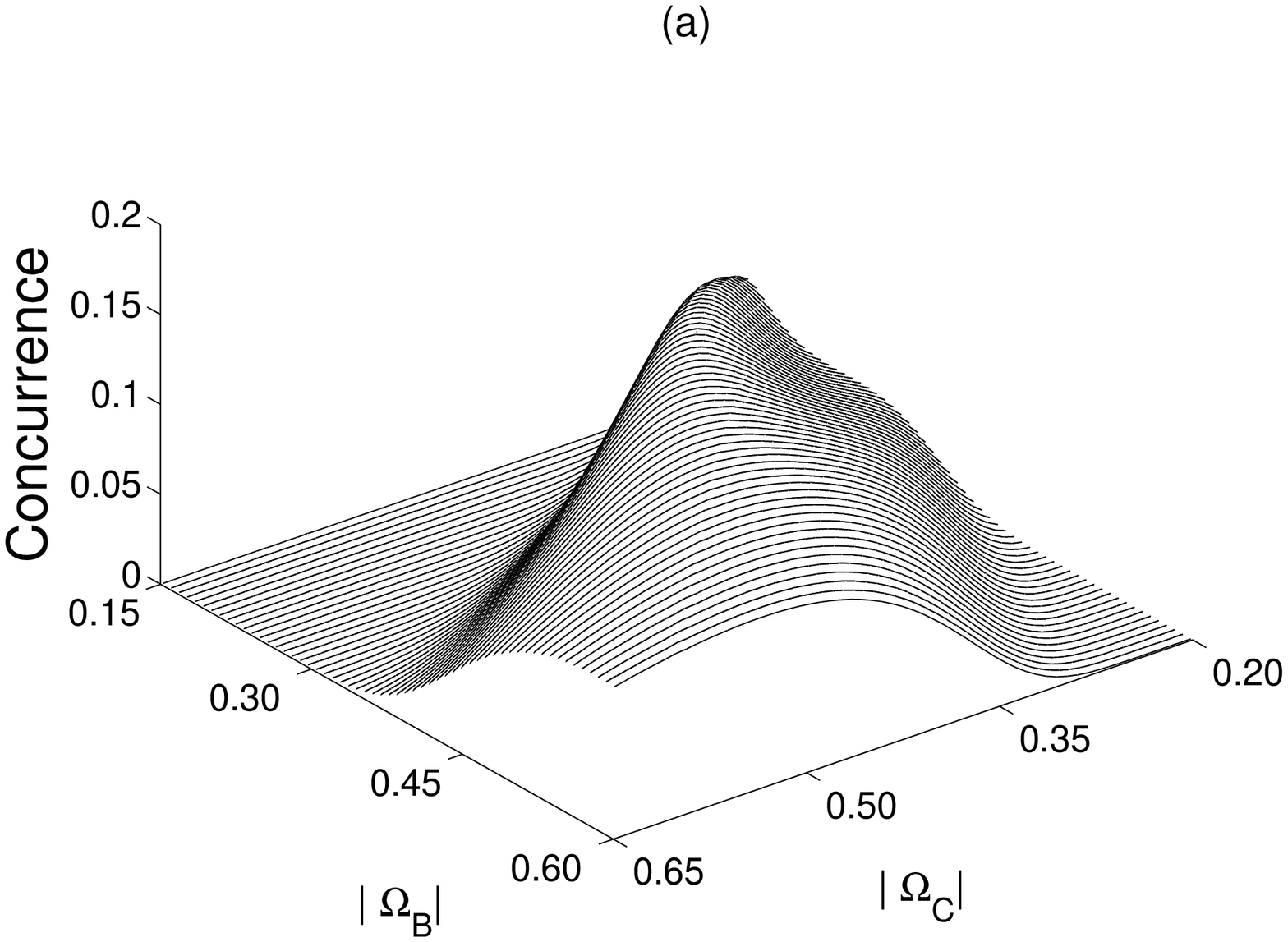}
\includegraphics[scale=0.4]{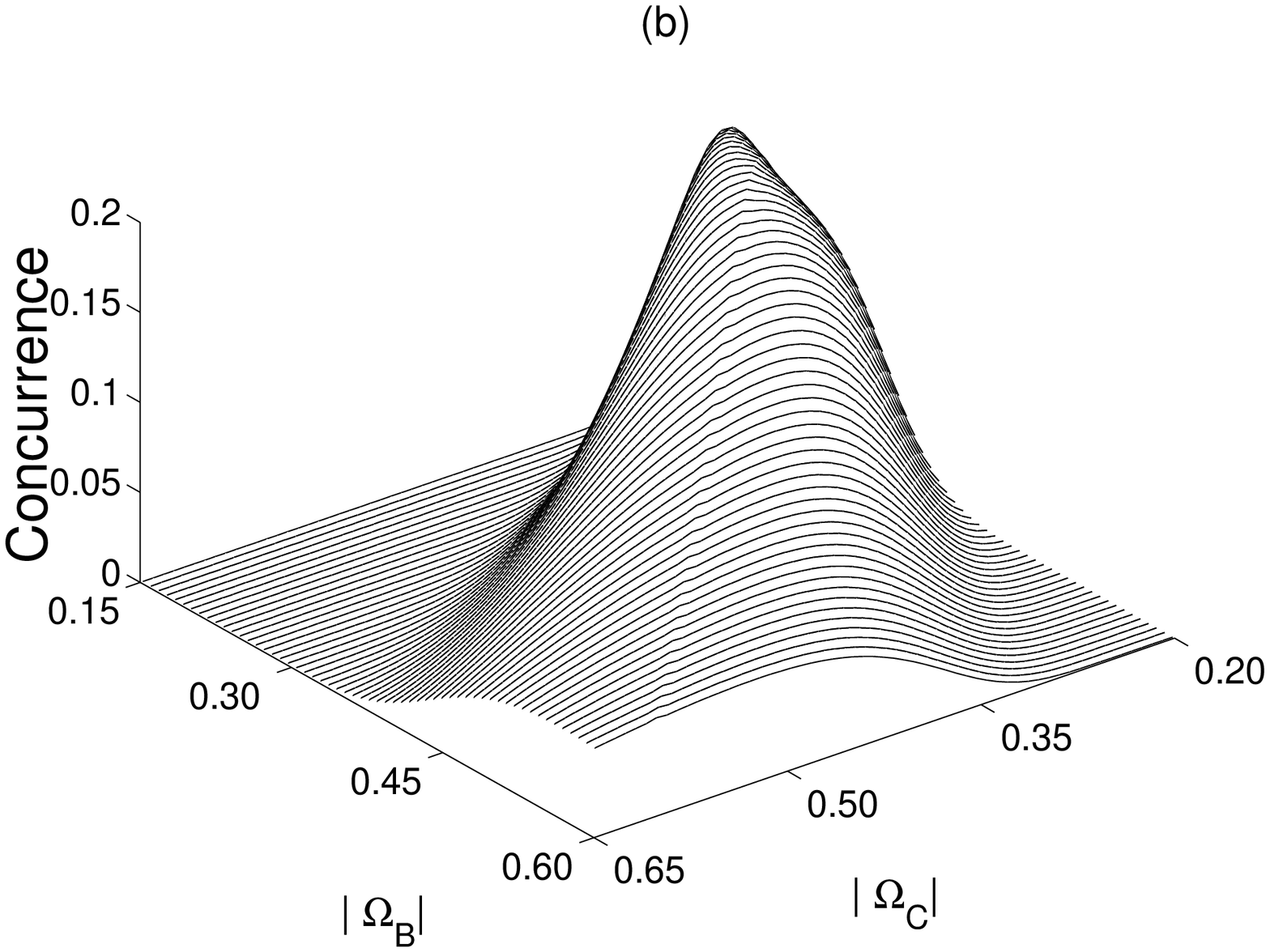}
\caption{\footnotesize The concurrence versus  Rabi frequencies of
two control fields using the first set of parameters in (a) and
the second set of parameters in (b). A large enhancement of
concurrence emerges when the group velocity matching condition is
approximately satisfied.}
\end{figure}

We have made a numerical computations on the concurrence versus
the Rabi frequencies of two control fields using the two sets of
parameters, given above (except for the control fields). The
results are shown in Fig. 2(a) and (b), respectively. The range of
frequency of the control fields are chosen as $1.5\gamma <
\Omega_{B}<6.0\gamma$ and $2.0\gamma<\Omega_{C}<6.5\gamma$. From
the results shown in Fig. 2(a) and Fig. 2(b)  we see that there is
a significant enhancement of concurrence for some values of the
control fields $\Omega_P$ and $\Omega_T$ when the group velocity
matching condition is approximately satisfied. Thus the
entanglement of the probe and trigger fields can  indeed  be
controlled in our system.

\begin{figure}
\centering
\includegraphics[scale=0.45]{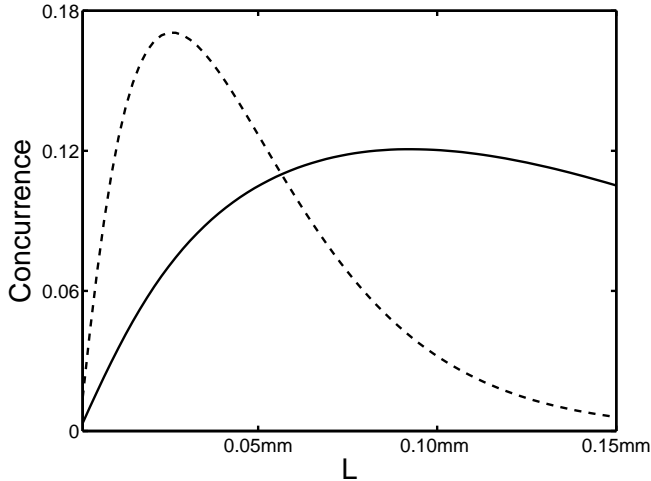}
\caption{\footnotesize The concurrence versus the medium length
 $L$ under the two sets of parameters (the same as
in Fig. 2). The solid line corresponds to the first set of
parameters while the dash line corresponds to the second ones.}
\end{figure}

In Fig. 3, we plot the concurrence versus the  medium length $L$
under the two sets of parameters (the same as those in Fig. 2(a)
and Fig. 2(b), respectively). The solid line corresponds to the
first set of parameters while the dash line corresponds to the
second ones. From the figure we find that although a long medium
permits a more sufficiently interaction between the two optical
pulses and favor to the entanglement, it also leads to larger
effects of absorption and dispersion, which impair the
entanglement when $L$ becomes large. Consequently, an appropriate
medium  length is necessary for obtaining the largest entanglement
of the probe and trigger fields.

\section{Conclusion}

We have proposed a scheme to create a pair of entangled photons by
using a novel five-level EIT configuration. We have shown that,
under suitable conditions, a large cross-phase modulation appears
when the group velocities of two pulses, the probe and the
trigger, are both small and comparable because of the symmetry of
the system. In our approach the effect of absorption and
dispersion of the system has been taken into account, which
contribute to the deformation of pulse shapes. The binary
information is encoded in the polarization degree of freedom of
the probe and the trigger pulses and the entanglement of the probe
and trigger pulses comes from the cross-Kerr nonlinearity. The
entanglement can be practically controlled by adjusting the
parameters of the system, such as the Rabi frequencies of two
coupling fields. The controllable entanglement suggested here may
facilitate promising applications for quantum information and
computation. The results presented in this work may be useful for
guiding an experimental finding of the photon-photon entanglement
in atomic systems.

\acknowledgments The authors thank Dr. Weiping Zhang and Dr.
Jinming Liu for useful discussions. The work was supported by the
Key Development Program for Basic Research of China under Grant
Nos. 2001CB309300 and 2005CB724508, and NSF-China under Grant Nos.
10434060 and 90403008.



\begin{references}
\bibitem{note1}When discussing the entanglement between the probe and
trigger fields, realistic parameters for these conditions have
been chosen, see Sec. IV.
\bibitem{note2}The effects of the absorption and dispersion on the
entanglement of the probe and trigger fields will be discussed in
the next section.
\end{references}
\end{document}